\documentclass[aps,pre,reprint,10pt,superscriptaddress,showpacs]{revtex4-1}
\usepackage{amsfonts} 
\usepackage{amsmath}
\usepackage{amssymb}
\usepackage{graphicx}
\usepackage{subfigure}
\usepackage{color}
\usepackage{color}
\usepackage{soul}
\usepackage{cancel}
\usepackage{ulem}
\newcommand*\xbar[1]{%
  \hbox{%
    \vbox{%
      \hrule height 0.5pt 
      \kern0.5ex
      \hbox{%
        \kern-0.1em
        \ensuremath{#1}%
        \kern-0.1em
      }%
    }%
  }%
} 
\begin{document}
\title{Neutrino magnetohydrodynamic instabilities in  presence of two-flavor oscillations}
\author{Debjani Chatterjee}
\email{chatterjee.debjani10@gmail.com}
\affiliation{Department of Applied Mathematics, University of Calcutta, Kolkata 700 009, India}
 \author{Amar P. Misra}
\email{apmisra@visva-bharati.ac.in; apmisra@gmail.com}
\affiliation{Department of Mathematics, Siksha Bhavana, Visva-Bharati University, Santiniketan-731 235,  India}
\author{Samiran Ghosh}
\email{ sran{\_}g@yahoo.com }
\affiliation{Department of Applied Mathematics, University of Calcutta, Kolkata 700 009, India}
\begin{abstract}
The influence of neutrino flavor oscillations on the propagation of magnetohydrodynamic (MHD) waves and instabilities is studied in neutrino-beam driven magnetoplasmas.   Using the neutrino MHD model, a general dispersion relation is derived which manifests  the  resonant interactions of MHD waves, not only with the neutrino beam,    but also  with the  neutrino  flavor oscillations.   It is found that the latter contribute  to   the wave dispersion and   enhance the magnitude of the instability of oblique magnetosonic waves. However, the shear-Alfv{\'e}n wave remains unaffected by the neutrino beam and neutrino flavor oscillations.    Such an enhancement of the magnitude of the instability of magnetosonic waves  can be  significant for relatively long-wavelength perturbations in the regimes of  high neutrino number density and/or strong  magnetic field,  giving a convincing mechanism  for type-II  core-collapse supernova explosion.  
 
\end{abstract}
\maketitle

\section{Introduction} \label{sec-introduct}
 Neutrinos  are generally produced due to very high explosions in the core of massive stars and   can have significant impact on the cooling of white dwarfs and neutron stars \cite{adams1963, winget2004}.
 The most apparent source of neutrinos is the Sun, where they are produced due to the simplest nuclear fusion reaction in which  two protons combine to form a deuterium nucleus with the emission of a positron and a neutrino. All other complex reaction processes that lead to heavier elements can also produce neutrinos which get away from the Sun at the speed of light in vacuum. Neutrinos are also produced by cosmic rays hitting up nuclei in the Earth's atmosphere, similar to the reactions of terrestrial high-energy particle accelerators. Such nuclear reactions  result not only in electron neutrinos as in the Sun, but also in two other flavors, namely, muon neutrinos and tau neutrinos--all of which were detected by super-Kamiokande detector \cite{meszaros2018}. On the other hand,   the neutrino-producing fusion reactions in stars do not release energy in the form of light or heat that could provide pressure to stop gravitational collapse of the stellar core. So, the collapse occurs and it  continues until the density in a nucleus is close to that in the core and suddenly a   massive explosion   occurs in producing all flavors of neutrinos. Such a sudden  and higher optical gleam   is known as a core-collapse supernova, e.g., SN1987A \cite{supernova1987a}.  
   Although the interaction between neutrinos and matter is weak, in the gamma-ray bursts of a supernova explosion, the energy emitted from neutrinos can be very high (almost $99 \%$ of the gravitational binding energy of   collapsing stars) and the intensity can be   more than $10^{28}$ W cm$^{-2}$. Furthermore, in the first few seconds of explosion, the neutrino burst  that originates  from the core of supernova is a source of free energy to drive collective oscillations and instabilities which may  lead  to the  revival of a stalled supernova shock \cite{bingham1994, bingham1996}.  
Typically, neutrinos produced in the solar atmosphere or in the core-collapse of stars   have energies ranging from  $1$ to $30$ MeV. However, recent observations with IceCube data
have indicated that  neutrinos can have energies  more or less $10^{15}$ eV  \cite{aartsen2014,meszaros2018}. Such high-energy neutrinos are expected to be produced in
astrophysical objects via the interactions of highly relativistic charged particles (Cosmic rays)    with either target particles or photons \cite{meszaros2018}.   
\par  
Neutrinos produced from different sources can play significant roles in the formation of  galaxies, galaxy clusters, and  various coherent structures at large scales. Apart from their possible gravitational interactions, neutrinos interact weakly with matter and thus are very important in astrophysics. In regions where other particles get trapped or move through slow diffusion processes, neutrinos can still escape from them and thus connect those of matter without being  detached from each other. In very hot or dense astrophysical objects, the emission of  neutrinos can be an important energy-loss mechanism. The energy transfer rate can be faster and very  efficient since neutrinos have almost zero mass and can travel at   relativistic speeds.  Furthermore, since neutrinos produced in the Sun can be detected at the Earth, they are useful to study nuclear   reactions that can occur in the core of massive stars. Also, because neutrinos are electrically neutral like photons and hence uninfluenced by the strong magnetic fields, they  tend to move back to the  creation regions, and thus can provide useful information about the regions where   particle creation and acceleration take  place in the Universe.   For more information about roles of neutrinos, see, e.g., \cite{role-neutrino}.   
\par 
In stellar environments,  the collective plasma effects can remarkably modify the production rate of neutrinos, e.g.,  the decay of photons and plasmons into neutrino pairs, which is the dominant neutrino emission mechanism at high-density plasmas.  The neutrino emission can also be possible in dense hot matter due to electron-positron annihilation \cite{amsterdamski1990}, in ultra-relativistic plasmas due to positron and plasmino annihilation \cite{braaten1992}.      
The neutrinos  interacting with plasmas  play key roles  in many astrophysical situations including  supernova explosions.  Such interactions not only reform the neutrino flavor oscillations (in which neutrinos oscillate from one flavor state to another) and initiate resonant interactions of different flavors \cite{bethe1986, mikheev1986,wolfenstein1978}, but also produce  an induced effective neutrino charge  as well as induced electric and magnetic fields that can lead to collective plasma oscillations and an enhancement of collision cross sections.  
 In this context, several authors have studied the neutrino-plasma interactions considering neutrino flavor oscillations, See, e.g.,  \cite{bingham2004, mendonca2014,haas2013,mendonca2013}. To discuss  a few, in Ref. \cite{haas2013}, it has been shown that the two-flavor  neutrino-plasma  oscillation equations admit an exact analytic solution for arbitrarily chosen electron neutrino populations. A hydrodynamic  model  has been introduced by Mendon{\c{c}}a and Haas  \cite{mendonca2013} to study  the plasma and  neutrino flavor oscillations in   turbulent plasmas.
\par  
In other contexts,  the neutrino-plasma coupling in magnetized plasmas can  lead to different types of hydrodynamic instabilities which may influence the neutrino beam transport by improving the properties of the background medium. Several studies have focused on the physics of collective neutrino-plasma interactions in different astrophysical situations \cite{bingham1996, serbeto1999, serbeto2002}. Also, the parametric instabilities in   intense neutrino flux and collective plasma oscillations have been studied by Bingham \textit{et al.}  \cite{bingham1994}. Furthermore, the generation of neutrino-beam driven wakefields \cite{aserbeto2002}, neutrino streaming instability  \cite{silva2000, silva1999,silva2006}, and neutrino Landau damping \cite{silva1999}    have   been studied in different contexts. The latter effect can be implemented to the cooling process of strongly turbulent plasmas. Furthermore,  it has   been shown that   neutrinos can contribute to the generation of both the inhomogeneities and magnetic fields in the early universe  \cite{shukla1998, shukla2003}. 
\par 
Recently, Haas \textit{et al.} \cite{haas2016}   proposed a neutrino MHD (NMHD) model in magnetoplasmas by  considering the neutrino-plasma interactions as well as the coupling between MHD waves and neutrino fluids. This model was studied for the propagation of magnetosonic waves in a specific geometry, i.e., when the propagation direction is perpendicular to the external magnetic field. However, the theory was later advanced with an arbitrary direction of propagation   \cite{haas2017}. Motivated by these works,  the influence of intense neutrino beams on the hydrodynamic Jeans instability has been studied by Prajapati  in a magnetized quantum plasma \cite{prajapati2017}. It turns out that the NMHD model has become very useful to establish connections between   various astrophysical phenomena and neutrino-plasma coupling processes in   magnetized media. 
\par 
In this work, we aim to advance the previous theory of NMHD waves \cite{haas2017} by considering (in addition to  the neutrino beam effects) the influence of two neutrino favor (electron- and muon-neutrinos)  oscillations on the neutrino-beam driven  MHD waves and instabilities. We show that the two-flavor oscillations, not only resonantly interact with the oblique magnetosonic wave, but can have a significant contribution to the growth rate of instability. 
 \par 
The  paper is organized as follows: In Sec. \ref{sec-model}, we   describe  the  NMHD model, which is coupled to the dynamics of two   neutrino flavors, namely, the  electron-neutrino and  muon-neutrino.  Using the perturbation analysis, a  general linear dispersion relation is derived in Sec. \ref{sec-linear} to show the coupling of MHD waves with the resonant neutrino beam and the  resonant neutrino flavor oscillations.  The instability growth rates for both the fast and slow magnetosonic waves are obtained  in Sec. \ref{sec-instab}, and analyzed numerically in Sec. \ref{sec-results}.         Finally, Sec. \ref{sec-conclu} is left for concluding remarks.
\section{Physical Model} \label{sec-model}
We consider a homogeneous magnetized system composed of   electrons and ions, as well as  the neutrino beams  of electron neutrinos and muon neutrinos. We also assume that the fluid descriptions for both the plasma electrons  and ions, and the neutrino beams are valid for the length scale  of the order of electron skin depth and the time scale  of the order of ion gyroperiod.         In the NMHD description, the continuity and momentum equations for the MHD fluids read    \cite{haas2017}
\begin{equation}
\frac{\partial \rho_m}{\partial t} + \nabla \cdot (\rho_m {\bf{U}})=0, \label{continuity-eq}
\end{equation}

\begin{equation}
\frac{\partial {\bf{U}}} {\partial t} +{\bf{U}} \cdot \nabla{\bf{U}}= -V_s^2 \frac{\nabla \rho_m}{\rho_m} + \frac{(\nabla \times {\bf{B}})\times {\bf{B}}}{\mu_0 \rho_m}+\frac{F_\nu}{m_i}, \label{momentum-eq}
\end{equation}
where $\rho_m= m_en_e+m_in_i\approx nm_i$ (with $n_e=n_i=n$) is the mass density,   ${\bf U}=(m_en_e{\bf u}_e+m_in_i{\bf u}_i)/(m_en_e+m_in_i)\approx (m_e{\bf u}_e+m_i{\bf u}_i)/m_i$ is the plasma velocity, $\mu_0$ is the permeability of free space, $V_s=\sqrt{k_BT_e/m_i}$ is the ion-acoustic velocity, and $F_{\nu}$ is the neutrino-plasma (electroweak) interaction force. Here,  $m_{e(i)}$ denotes the electron (ion) mass, $n_{e(i)}$   the electron (ion) number density,   $\mathbf{u}_{e(i)}$   the electron (ion) fluid velocity, ${\bf B}$ is the magnetic field,   $T_e$ is the electron temperature, and $k_B$ the Boltzmann constant. In addition,   the equation for the magnetic flux modified by the electroweak force is given by
\begin{equation}
\frac{\partial {\bf B}}{\partial t} = \nabla \times \left( {\bf U} \times {\bf B} -\frac{F_\nu}{e}\right), \label{B-eq}
\end{equation}
where $e$ is the elementary charge, $F_\nu=\sqrt{2} G_F ({\bf E}_\nu + {\bf U} \times {\bf B}_\nu)$ with $G_F$ denoting the Fermi  coupling constant and $E_{\nu}~(B_{\nu})$  the effective  electric  (magnetic) field induced by the weak interactions of neutrinos with plasmas, given by,
\begin{eqnarray}
&&{\bf E}_\nu= -\nabla N_e -\frac{1}{c^2} \frac{\partial}{\partial t} (N_e {\bf v}_e),\\
&&{\bf B}_\nu= \frac{1}{c^2} \nabla \times (N_e {\bf v}_e).
\end{eqnarray}
\par 
Here,   $N_e~({\bf v}_e)$ denotes the number density (velocity) of electron neutrinos. For a coherent neutrino beam with an  energy ${\cal E}_0$,  the continuity equations for electron    and muon neutrinos (with number density $N_\mu$, velocity $\mathbf{v}_\mu$), respectively, are \cite{mendonca2014,haas2019}
 \begin{equation}
\frac{\partial N_e}{\partial t} + \nabla \cdot (N_e {\bf v}_e) =\frac{1}{2} N \Omega_0 P_2, \label{electron-neutrino-continuity-eq}
\end{equation}
\begin{equation}
\frac{\partial N_\mu}{\partial t} + \nabla \cdot (N_\mu {\bf v}_\mu) =-\frac{1}{2} N \Omega_0 P_2, \label{muon-neutrino-continuity-eq}
\end{equation}
where $P_2$ corresponds to the neutrino coherence  in the flavor polarization vector ${\bf P}=(P_1, P_2, P_3)$, $N=N_e+N_\mu$ is the total neutrino fluid density, and $\Omega_0=\omega_0 \sin{(2\theta_0)}$. Here,  $\omega_0= \delta{m^2} c^4/2 \hbar {\cal E}_0$ with $\delta{m^2}$ denoting the squared neutrino mass difference, $c$ the speed of light in vacuum, $\hbar$ the reduced Planck's constant,      and $\theta_0$  the neutrino oscillation  mixing angle. While the left-hand sides of Eqs. \eqref{electron-neutrino-continuity-eq} and \eqref{muon-neutrino-continuity-eq} involve the convective terms due to the flows of neutrinos into plasmas,  the terms on the right-hand sides appear due to the neutrino flavor oscillations along with the rates of changes of  the electron- and muon-neutrino fluid densities. We also require  the global neutrino fluid densities to be  conserved, i.e., 
\begin{equation}
\frac{d}{dt} \int (N_e+N_\mu) d^3 {\bf r}= - \int \nabla \cdot (N_e {\bf v}_e + N_\mu {\bf v}_\mu)d^3 {\bf r}=0. \label{neutrino-preserve-eq}
\end{equation}
Next, the electron neutrino and muon neutrino equations of motion are
\begin{equation}
\frac{\partial {\bf p}_e}{\partial t} + {\bf v}_e \cdot \nabla {\bf p}_e = -\frac{\sqrt{2} G_F}{m_i} \nabla \rho_m, \label{electro-neutrino-force-eq}
\end{equation}
\begin{equation}
\frac{\partial {\bf p}_\mu}{\partial t} + {\bf v}_\mu \cdot \nabla {\bf p}_\mu =0, \label{muon-neutrino-force-eq}
\end{equation}
 where ${\bf p}_e ={\cal{E}}_e {\bf v}_e/c^2$ and ${\bf p}_\mu ={\cal{E}}_\mu {\bf v}_\mu/c^2$ are the   momenta of electron and muon neutrinos with ${\cal{E}}_{e,\mu}=  \left(p_{e,\mu}^2c^2+m^2_{e,\mu}c^4\right)^{1/2}$ denoting the   electron- and muon-neutrino energies, $m_{e,\mu}$   the electron (muon) neutrino mass,  and $v_{e(\mu)}$ the electron (muon) neutrino velocity.
 \par
To complete the description of neutrino-plasma interactions, we require the time evolution equations of the components of the flavor polarization vector ${\bf P}=(P_1, P_2, P_3)$ as \cite{mendonca2014} 
\begin{equation}
\frac{d P_1}{d t}=-\Omega(n_e)P_2, \label{P1-eq}
\end{equation}
\begin{equation}
\frac{d P_2}{d t}=\Omega(n_e)P_1 -\Omega_0 P_3, \label{P2-eq}
\end{equation}
\begin{equation}
\frac{d P_3}{d t}=\Omega_0 P_2, \label{P3-eq}
\end{equation}
where $\Omega(n_e)=\omega_0[\cos (2\theta_0) -\sqrt{2} G_F n_e/(\hbar \omega_0)]$. The total time derivatives appearing in Eqs. \eqref{P1-eq}-\eqref{P3-eq} should, in general, be different. However, for a mono-energetic neutrino beam, the velocity of each neutrino flavor can be assumed to be identical so that $\mathbf{v}_e=\mathbf{v}_\mu=\mathbf{v}$. One can then consider the total time derivative  as $d/dt\equiv \partial_t+\mathbf{v}\cdot\nabla$. Since we are interested in the linear regime, the convective parts will be less important and can thus be disregarded in the analysis in Sec. \ref{sec-linear}.    
\section{Linear waves: General dispersion relation} \label{sec-linear}
In order to obtain a general dispersion relation for NMHD waves, we Fourier analyze the system of Eqs. \eqref{continuity-eq}-\eqref{P3-eq} about the following equilibrium state:
\begin{equation}\label{eqbm}
\begin{split}
 &{\bf U}=0,~N_e=N_{e0},~N_\mu=N_{\mu 0},\\
 &{\bf v}_e= {\bf v}_\mu ={\bf v}_0, ~N_0=N_{e0}+N_{\mu 0},\\ 
&P_1=\frac{\Omega_0}{\Omega_\nu},~ P_2=0,~ P_3=\frac{\Omega(n_0)}{\Omega_\nu}=\frac{N_{e0}-N_{\mu 0}}{N_0},
\end{split}
\end{equation}
where  $\Omega_\nu =\sqrt{\Omega^2(n_0)+\Omega_0^2}$ is the eigenfrequency of two-flavor neutrino oscillations and $n_0$ is the background number density of electrons and ions. 
We mention that  $P_2(0)=0$ is considered in Eq. \eqref{eqbm} without any loss of generality. The reason is that Eqs. \eqref{P1-eq}-\eqref{P3-eq}   can be reduced to an equation for $P_2$ \cite{mendonca2014}, i.e., 
\begin{equation}\label{P22}
\frac{d^2P_2}{dt^2}+\bar{\omega}^2P_2=0~\rm{with}~\bar{\omega}^2=\Omega^2+\Omega_0^2,
\end{equation}
whose solution can be obtained in the form  $P_2(t)=A\sin{\bar{\omega}t}$ for some constant $A$ and assuming the  phase constant as zero. Also, $P_1(0)$,  $P_2(0)$, and $P_3(0)$ are such that $|P(0)|=1$.
\par 
Next,  assuming the  MHD perturbations in the form of plane waves $\sim \exp [i({\bf k} \cdot {\bf r} -\omega t)]$ with wave vector $\mathbf{k}$ and wave frequency $\omega$, we obtain from Eqs.   \eqref{continuity-eq} to \eqref{B-eq} the following expression for the perturbed velocity.
\begin{eqnarray}
&&\omega^2 \delta{\bf U} =(V_s^2+V_A^2)({\bf k} \cdot \delta{\bf U}) {\bf k} + ({\bf k} \cdot {\bf V}_A) \left\lbrace ({\bf k} \cdot {\bf V}_A) \delta{\bf U} \right. \notag\\
&& \left. - (\delta{\bf U} \cdot {\bf V}_A) {\bf k} -({\bf k} \cdot \delta{\bf U}) {\bf V}_A \right\rbrace + \frac{\sqrt{2} G_F}{m_i c^2} \omega \left[c^2 {\bf k} -\omega {\bf v}_0) \delta N_{e} \right. \notag\\
&& \left. -\omega N_{e0} \delta{\bf v}_{e}\right],\label{dispersion}
\end{eqnarray}
where $\delta{f}$ denotes the perturbation of a physical quantity $f$.  Also, from Eq. \eqref{electro-neutrino-force-eq}, we have 
\begin{multline}
 (\omega- {\bf k} \cdot {\bf v}_0) \delta {\bf p}_{e} = \left[\delta{\bf v}_{e} +\left(1-\frac{v_0^2}{c^2} \right)^{-1} \frac{{\bf v}_0 \cdot \delta{\bf v}_{e} }{c^2} {\bf v}_0\right]  \\
  = \sqrt{2} G_F \frac{k}{m_i} \delta \rho_{m}.  
\end{multline}
So, for nonrelativistic fluid flow with $v_0\ll c$, we get
\begin{multline}
\delta {\bf v}_{e} = \frac{\sqrt{2} G_F}{{\cal{E}}_0 (\omega-{\bf k} \cdot {\bf v}_0)} \left[  \frac{c^2 {\bf k} \delta \rho_{m}}{m_i} \right.\\
\left.-\left( \frac{{\bf k} \cdot {\bf v}_0 \rho_{m1}}{m_i} -\frac{n_0 \omega}{c^2} {\bf v}_0 \cdot \delta {\bf U}\right) \right]. \label{ve-eq}
\end{multline}
 Using the perturbed form of Eq. \eqref{continuity-eq}, we can rewrite Eq. \eqref{ve-eq} as
 \begin{equation}
 \delta{\bf v}_{e} = \frac{\sqrt{2} G_F}{{\cal{E}}_0 (\omega-{\bf k} \cdot {\bf v}_0)} \frac{c^2 \rho_{m0}}{m_i \omega} ({\bf k} \cdot \delta{\bf U}) {\bf k}. \label{ve-final-eq}
 \end{equation}
\par
Next, from Eqs. \eqref{P1-eq} - \eqref{P3-eq} one obtains
\begin{equation}
\delta P_2= -i \frac{\sqrt{2} \Omega_0 \omega G_F}{(\omega^2-\Omega_\nu^2) m_i \hbar \Omega_\nu} \delta\rho_{m}, \label{derived-P2-eq}
\end{equation} 
and using this expression of $\delta P_2$, we obtain    from Eq. \eqref{electron-neutrino-continuity-eq}   the following equation for the perturbed density.
\begin{eqnarray}
&& \delta N_{e}=N_{e0} \frac{\sqrt{2}G_F c^2 \rho_{m0}}{{\cal{E}}_0 (\omega- {\bf k} \cdot {\bf v}_0)^2 m_i \omega} k^2 ({\bf k} \cdot \delta{\bf U}) \notag\\
&& +\frac{\sqrt{2} G_F \Omega_0^2 N_0 \rho_{m0}}{2 m_i \hbar \Omega_\nu (\omega- {\bf k} \cdot {\bf v}_0)(\omega^2-\Omega_\nu^2)} ({\bf k} \cdot \delta {\bf U}). \label{derived-Ne2-eq}
\end{eqnarray}
\par 
The expressions for the perturbed density and velocity of electron neutrinos [Eqs. \eqref{ve-final-eq} and \eqref{derived-Ne2-eq}] together with the continuity equation \eqref{electron-neutrino-continuity-eq} can then be used in  the expression for the perturbed neutrino fluid force to show that its magnitude is enhanced for $\omega\approx {\bf k} \cdot {\bf v}_0$ and/or $\omega\approx \Omega_\nu$. It follows that   the MHD waves can have resonant-like  interactions with the streaming neutrino beam and the neutrino flavor oscillations for which the  energy exchange can take place  leading to the MHD instability.  Although the resonant contribution of the neutrino beam is known in Ref. \cite{haas2017}, we will, however, consider both the resonances in order to study the relative influence of the neutrino flavor oscillations on the MHD instability.   
Finally, from  Eqs.  \eqref{dispersion}, \eqref{ve-final-eq}, and \eqref{derived-Ne2-eq}, we obtain  the following  dispersion relation.
\begin{eqnarray}
&& \omega^2 \delta {\bf U}= \left\lbrace V_s^2+ V_A^2+V_N^2 \frac{c^2 k^2 -\omega^2}{(\omega- {\bf k} \cdot {\bf v}_0)^2} \right\rbrace ({\bf k} \cdot \delta {\bf U}) {\bf k} \notag\\
&& +({\bf k} \cdot  {\bf V}_A)\left\lbrace ({\bf k} \cdot  {\bf V}_A) \delta {\bf U} -(\delta {\bf U} \cdot  {\bf V}_A) {\bf k} - ({\bf k} \cdot \delta {\bf U}) {\bf V}_A\right\rbrace \notag\\
&& +V_\text{osc}^2 \frac{\Omega_0^2 \omega {\cal{E}}_0(c^2k^2 - \omega ({\bf k} \cdot  {\bf v}_0))}{2c^2k^2 \hbar \Omega_\nu(\omega-{\bf k} \cdot  {\bf v}_0)(\omega^2-\Omega_\nu^2)} {\bf k} (\bf k \cdot \delta {\bf U}). \label{dispersion-relation}
\end{eqnarray}
Here, ${V}_A= { B}_0/(\mu_0 \rho_{m0})^{1/2}$ is the Alfv{\'e}n velocity associated with the MHD wave, $V_N = \left[2 G_F^2 \rho_{m0} N_{e0}/(m_i^2 {\cal{E}}_0)\right]^{1/2}$ is the velocity associated with the  electron-neutrino beam (which involves the densities of both the MHD   and   neutrino fluids, highlighting the mutual coupling between them), and $V_\text{osc} =\left[2 G_F^2 \rho_{m0} N_0/(m_i^2 {\cal{E}}_0)\right]^{1/2}$ that due to the coupling of MHD waves with  both the electron- and muon-neutrino flavor oscillations. Thus,  the terms proportional to $V_N^2$ and $V_\text{osc}^2$ in  Eq. \eqref{dispersion-relation} appear due to the neutrino beam effect (electron-neutrino) and two-flavor (both the electron-neutrino and muon-neutrino)   oscillations.  
As noted before and   is clear from Eq. \eqref{dispersion-relation} that, in addition to the phase velocity resonance (at the  neutrino beam velocity, i.e., $\omega\approx {\bf k} \cdot  {\bf v}_0$), there  also can occur  the resonance (at the frequency of two-flavor oscillations, i.e., $\omega\approx\Omega_{\nu}$) due to the coupling between  MHD waves and two neutrino flavor oscillations. Furthermore,  disregarding the contribution of the neutrino flavor oscillations from Eq. \eqref{dispersion-relation}, one can recover the same dispersion relation as in Ref. \cite{haas2017}. Thus, the dispersion equation  \eqref{dispersion-relation} generalizes the previous theory with the effects of neutrino flavor oscillations.  
\par 
We note that the adiabatic sound speed is also modified by the effects of neutrino flavor oscillations. Thus, defining 
$\tilde{V}_s^2$ by
\begin{eqnarray}
&&\tilde{V}_s^2(\omega,{\bf k}) =V_s^2 + V_N^2 \frac{c^2 k^2 -\omega^2}{(\omega- {\bf k} \cdot {\bf v}_0)^2} \notag\\
&& +V_{osc}^2 \frac{\Omega_0^2 \omega {\cal{E}}_0(c^2k^2 - \omega ({\bf k} \cdot  {\bf v}_0))}{2c^2k^2 \hbar \Omega_\nu(\omega-{\bf k} \cdot  {\bf v}_0)(\omega^2-\Omega_\nu^2)}, \label{VS-eq}
\end{eqnarray}
Eq. \eqref{dispersion-relation} can be recast as
\begin{eqnarray}
&&\omega^2 \delta {\bf U}= (V_A^2+\tilde{V}_s^2) ({\bf k} \cdot \delta {\bf U}) {\bf k} +({\bf k} \cdot  {\bf V}_A)\left\lbrace ({\bf k} \cdot  {\bf V}_A) \delta {\bf U} \right. \notag\\
&& \left. -(\delta {\bf U} \cdot  {\bf V}_A) {\bf k} - ({\bf k} \cdot \delta {\bf U}) {\bf V}_A\right\rbrace. \label{dispersion-reduced}
\end{eqnarray}
The form of Eq. \eqref{dispersion-reduced} is the same as the well-known dispersion equation for the propagation of linear waves in a compressible, nonviscous, perfectly conducting magnetofluid. Consequently, the usual method of recovering various modes applies, which we discuss as follows.  
\par 
We consider the wave propagation at an arbitrary angle $\theta$ with respect to the constant magnetic field ${\bf B}_0=B_0\hat{z}$  and assume, without loss of generality,   that  the wave vector ${\bf k}$  lies in the $xz$-plane. Thus,  equating the coefficient determinant of the homogeneous system \eqref{dispersion-reduced} for the components of $\delta {\bf U}$ to zero, we obtain the following linear dispersion relation for the coupling of MHD waves with the neutrino beam and the neutrino flavor oscillations. 
\begin{eqnarray}
&& (\omega^2-k^2 V_A^2 \cos^2\theta)\left[\omega^4 -k^2(V_A^2+\tilde{V}_s^2 )\omega^2 \right. \notag\\
&& \left. +k^4 V_A^2 \tilde{V}_s^2 \cos^2\theta\right]=0. \label{final-dispersion-relation}
\end{eqnarray}
From Eq. \eqref{final-dispersion-relation}, it is evident that the first factor, when equated to zero, gives  the dispersion relation for oblique Alfv{\'e}n or shear-Alfv{\'e}n waves, i.e.,    $\omega=k V_A \cos\theta$. Such waves are  neither influenced by the neutrino beam nor by the neutrino flavor oscillations. This is expected as shear-Alfv{\'e}n waves are characterized by both $\delta{\bf U}\cdot{\bf B_0}=0$ and ${\bf k}\cdot\delta{\bf U}=0$. The latter, however, eliminates the contributions of neutrinos in Eq. \eqref{dispersion-relation}.   So, we are interested in the second factor to obtain the following dispersion relation.
\begin{equation}
 \omega^4 -k^2(V_A^2+\tilde{V}_s^2 )\omega^2+k^4 V_A^2 \tilde{V}_s^2 \cos^2\theta=0. \label{eq-disp}
\end{equation}
Equation \eqref{eq-disp} reveals the coupling of the oblique magnetosonic waves with the neutrino beam and the neutrino two-flavor oscillations.  Note that for wave propagation perpendicular to the magnetic field $(\theta=\pi/2)$, typical magnetosonic mode is recovered \cite{haas2016}, which is, however, modified by the influence of neutrino flavor oscillations mediated through the term proportional to $V_\text{osc}^2$. Furthermore, due to smallness of the Fermi constant $G_F$, and hence $V_N^2$ and $V_\text{osc}^2$, the contributions from the neutrino beam and neutrino flavor oscillations are typically  small. So,  they   can be considered as small perturbations to the squared acoustic speed $V_s^2$. Physically, these perturbations, as they develop  in the resonant interactions of MHD waves with the streaming neutrino beam and the neutrino flavor oscillations, may lead to instabilities due to energy gain from neutrinos that can be radiated due to core collapse of massive stars in supernova explosions.     In Sec. \ref{sec-instab}, we will investigate the qualitative features of these instabilities in details.  

\section{Growth rate  of instability} \label{sec-instab}
 To study the instabilities of oblique magnetosonic waves, we rewrite the dispersion equation \eqref{eq-disp}   as
 \begin{eqnarray}
 &&\omega^4 -k^2(V_A^2+ V_s^2)\omega^2 +k^4V_A^2 V_S^2 \cos^2\theta \notag\\
 &&= V_N^2 k^2 \frac{\left(c^2 k^2 -\left({\bf k} \cdot  {\bf v}_0\right)^2\right)\left(\omega^2 -k^2V_A^2 \cos^2\theta\right)}{\left(\omega- {\bf k} \cdot {\bf v}_0\right)^2}\notag\\
 && +V_{osc}^2 \frac{\Omega_0^2 \omega {\cal{E}}_0\left(c^2k^2 - \omega \left({\bf k} \cdot  {\bf v}_0\right)\right)\left(\omega^2 -k^2V_A^2 \cos^2\theta\right)}{2c^2k^2 \hbar \Omega_\nu\left(\omega-{\bf k} \cdot  {\bf v}_0\right)\left(\omega^2-\Omega_\nu^2\right)}. \label{neutrino-dispersion}
 \end{eqnarray}
Since the influences of the neutrino streaming beam and the flavor oscillations on the instability growth rates are of our prime interest, we assume  
 \begin{equation}
 \omega=\tilde{\Omega} +\delta \omega,~~\rm{with}~|\delta \omega|\ll\tilde{\Omega},   \label{omega-eq}
\end{equation}
together with the double resonance condition 
\begin{equation}
 \omega=\Omega_\nu \approx\tilde{\Omega}={\bf k}\cdot {\bf v}_0, \label{eq-reso}
\end{equation}
where $\tilde{\Omega}$ is a  solution of the following dispersion equation (in absence of the effects of neutrinos)
\begin{equation}
\omega^4 -k^2(V_A^2+ V_s^2)\omega^2 +k^4V_A^2 V_S^2 \cos^2\theta=0. \label{omega-new-eq}
\end{equation}
From Eq. \eqref{omega-new-eq},  the frequencies of the fast (with the suffix $+$) and slow  (with the suffix $-$) classical magnetosonic modes can be obtained as
\begin{equation}
\omega= \tilde{\Omega}_{\pm} =k V_{\pm}, \label{eq-Omegapm}
\end{equation}
where $V_\pm$ are the corresponding phase velocities, given by,
\begin{equation}
V_\pm= \left[ \frac{1}{2}\left( V_A^2+V_s^2 \pm \sqrt{(V_A^2-V_s^2)^2 +4V_A^2 V_s^2 \sin^2 \theta}\right) \right] ^{1/2}. \label{V-eq}
\end{equation}
Thus, from Eqs.   \eqref{neutrino-dispersion} to \eqref{eq-reso}  and using the fact that $V_\pm\ll c^2$ for non-relativistic fluid flow,  we obtain
\begin{eqnarray}
&&(\delta \omega)^3 \approx \pm\left[ \frac{V_N^2 c^2 k^3 (V_\pm^2 -V_A^2 \cos^2 \theta)}{2 V_\pm \sqrt{(V_A^2-V_s^2)^2 +4V_A^2 V_s^2 \sin^2 \theta}} \right. \notag\\
&& \left. +\frac{G_F^2 \rho_{m0} N_0 \Omega_0^2(V_\pm^2 -V_A^2 \cos^2 \theta)}{4 V_\pm^2 \hbar m_i^2 \sqrt{(V_A^2-V_s^2)^2 +4V_A^2 V_s^2 \sin^2 \theta}}\right]. \label{delta-omega-eq}
\end{eqnarray}
The instability  growth rate $\gamma= \Im(\delta \omega)>0$ is then obtained as
\begin{equation}
\gamma\equiv \gamma_\pm=\left[\left(\gamma^{\pm}_{\nu}\right)^3+\left(\gamma^{\pm}_\text{osc}\right)^3\right]^{1/3},
\end{equation}
where we have defined the  dimensionless parameter $\Delta=V_N^2/c^2$  and the expressions for $\gamma^{\pm}_{\nu}$ and $\gamma^{\pm}_\text{osc}$, respectively,  are  
 \begin{multline}
   \gamma^{\pm}_{\nu}=\frac{\sqrt{3} k}{2^{4/3}} \left[ \frac{\Delta c^4 |V_\pm^2-V_A^2 \cos^2\theta |}{ V_\pm \sqrt{(V_A^2-V_s^2)^2 +4V_A^2 V_s^2 \sin^2 \theta}}\right] ^{1/3},  \\
   \gamma^{\pm}_\text{osc}=\frac{\sqrt{3} }{2^{4/3}}\left[\frac{G_F^2 \rho_{m0} N_0 \Omega_0^2|V_\pm^2-V_A^2 \cos^2\theta |}{2 V_\pm^2 \hbar m_i^2 \sqrt{(V_A^2-V_s^2)^2 +4V_A^2 V_s^2 \sin^2 \theta}}\right]^{1/3}. \label{eq-gamma}
 \end{multline}
Here, we again note that while  the quantity $\gamma^{\pm}_{\nu}$ is associated with the interactions of MHD waves with the streaming (with velocity $\mathbf{v}_0$) neutrino beam, the quantity $\gamma^{\pm}_\text{osc}$ appears due to coupling of MHD waves with neutrino two-flavor oscillations (with frequency $\Omega_\nu$). In absence of the latter, one can recover  exactly the same result as in Ref. \cite{haas2017}.  Furthermore, the conditions for the weak perturbations due to the neutrino beam and neutrino flavor oscillations can be validated so that $\gamma^{\pm}_{\nu}/\Omega_\nu\ll1$ and $\gamma^{\pm}_\text{osc}/\Omega_\nu\ll1$ since the terms in the square brackets in Eq. \eqref{eq-gamma} can be made less than unity after an appropriate normalization. 
\par 
The relative influence of the neutrino flavor oscillations on the growth rate of instabilities can be noted and it is given by
\begin{equation}
\frac{\gamma^{\pm}_\text{osc}}{\gamma^{\pm}_{\nu}}=\frac{1}{2^{4/3}k}\left[\frac{ \left(\delta{m^2}c^3\right)^2 \sin^2(2\theta_0)}{\hbar^3{\cal E}_0  V_\pm }\right]^{1/3}\left(\frac{N_0}{N_{e0}}\right)^{1/3}. \label{eq-ratio}
\end{equation}
From Eq. \eqref{eq-ratio}, it is to be noted that while the quantity $\gamma^{\pm}_{\nu}$ explicitly depends  on the wave number $k$, $\gamma^{\pm}_\text{osc}$ is independent of $k$, which means the growth rate  ratio is inversely proportional to $k$.  Thus, it follows that in the regimes of sufficiently large wave numbers (provided that the wavelength is not too small) of magnetosonic perturbations, the neutrino beam contribution to the MHD instability can be larger than that of neutrino flavor oscillations.  In contrast, the two-flavor oscillations can dominate over the neutrino beam-plasma interactions if initially the muon-neutrino beam density $(N_{\mu0})$ is much higher than that of electron neutrinos $(N_{e0})$ or the streaming neutrino spinor energy is relatively low. Furthermore, depending on the angle of propagation $\theta$, a relatively low  magnetic field strength and/or low  thermal energies of MHD fluids can enhance the neutrino flavor oscillation correction.
  
%
  \begin{figure*}[ht]
\centering
\includegraphics[height=2.8in,width=6.5in]{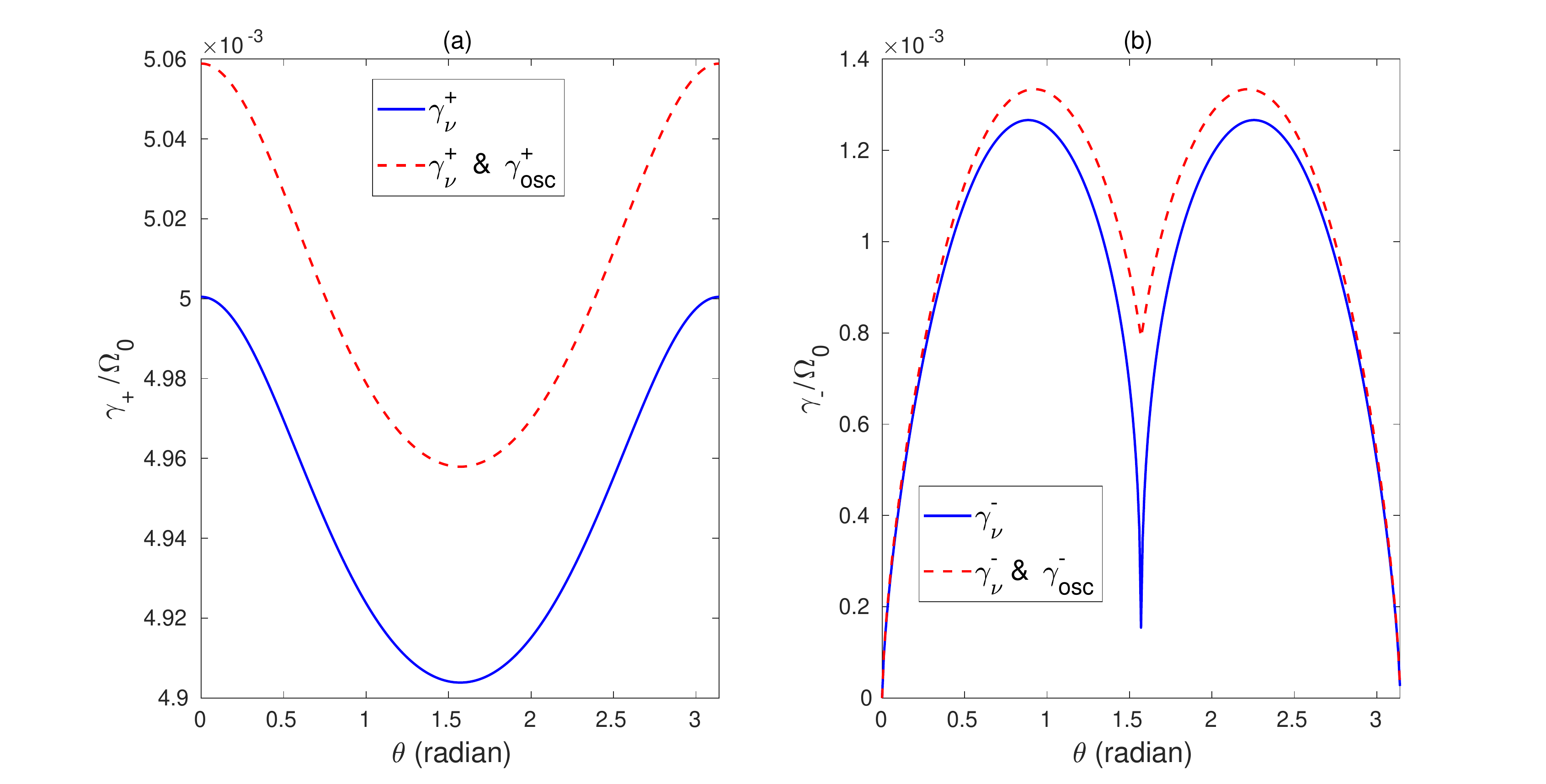}
\includegraphics[height=2.8in,width=6.5in]{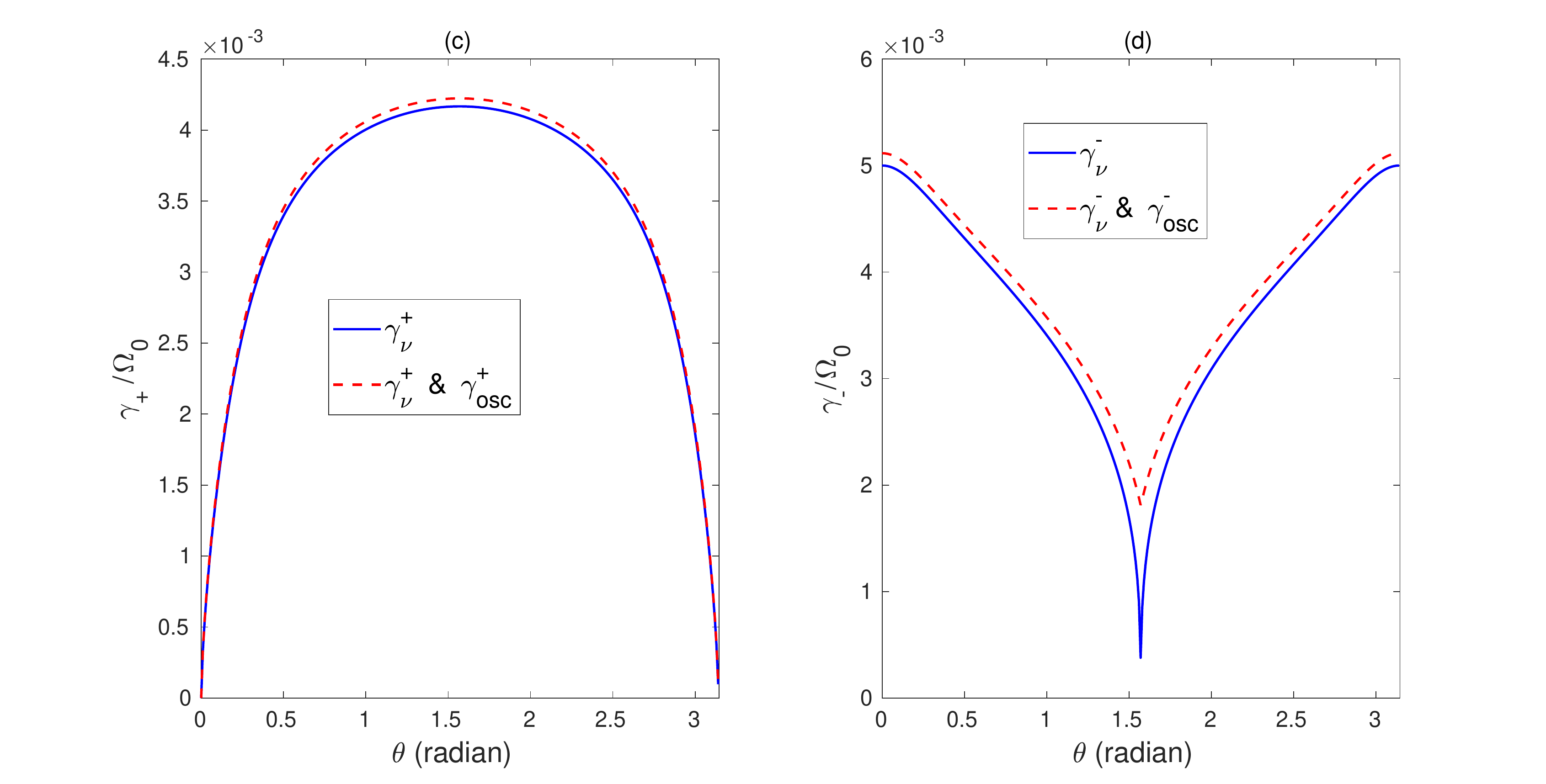}
\caption{ Instability growth rates of the fast (with $+$)  and slow (with $-$) oblique magnetosonic waves are shown against the    propagation angle $\theta$ with two different magnetic field strengths: (i) $B_0=5\times10^6$ T [Subplots (a) and (b)] and (ii)   $B_0=2\times10^7$ T  [Subplots (c) and (d)].  The solid and dashed lines correspond to the growth rates when only the neutrino beam effect is present and when both the neutrino beam and neutrino flavor oscillation effects are present. The value  $B_0=2\times10^7$ T is the critical value at which the  growth rates  in the upper  subplots  exhibit  (almost) an opposite trend.    }
\label{fig1}
\end{figure*}
\section{Results and discussion} \label{sec-results}
In order to examine the qualitative features, we numerically study the growth rates of instability for both the fast and slow magnetosonic waves, as well as the relative influence of the flavor oscillations. To this end, we consider the parameters that are relevant for type-II core-collapse supernova SN1987A \cite{haas2017}.     In such scenarios, we can expect a fluid flow of $10^{58}$ neutrinos of all flavors and the streaming  energy   ${\cal E}_0\sim10-15$ MeV. There may also exist  a strong magnetic field $B_0\sim 10^6 - 10^8$ T and high neutrino beam densities,  $N_0\sim10^{34}-10^{37}$ m$^{-3}$. We, however,   consider  $n_0=10^{34}$ m$^{-3}$, $N_{e0}=10^{37}$ m$^{-3}$, and  two different values of each of $N_0$ and $B_0$, namely $N_0=5\times10^{37}$ m$^{-3}$, $10^{38}$ m$^{-3}$ and  $B_0=5 \times 10^6$ T, $2 \times 10^7$ T. Furthermore,   $T_e=0.1$ MeV, $k=10^2$ m$^{-1}$ $\Delta m^2 c^4=3\times 10^{-6}~ \text{(eV)}^2$, $\sin (2\theta_0)=10^{-1}$, ${\cal{E}}_0=10 MeV$,  and $G_F=1.45\times10^{-62}$ J M$^3$. With these parameters,  the non-relativistic conditions $V_A/c\ll1$ and $V_s/c\ll1$, as well as   the simplifying assumption \cite{haas2016}: $ck/\omega_{pe}\ll\omega_{pe}/\omega_{ce}$, where $\omega_{pe}$ and $\omega_{ce}$ are, respectively,  the electron plasma and electron cyclotron frequencies,    for the present model    are also satisfied.
\par
 Figure \ref{fig1} displays the growth rates of both the fast and slow magnetosonic modes when only the influence of the neutrino beam is present  and when both the neutrino beam and the neutrino flavor oscillations are present. These growth rates are also plotted with two different magnetic field strengths. It is found that in the regimes of relatively low  magnetic field [Subplots (a) and (b)], the contribution  of the two-flavor oscillation  to the growth rate becomes significant. It enhances the growth rates of both the fast and slow magnetosonic modes. In this case, the growth rate for the fast mode exhibits an inverted bell-shaped curve, while that for the slow mode   displays a symmetric double-hump even in absence of the flavor oscillations.   Consequently,  the growth rate  of the fast mode  reaches its minimum at $\theta=\pi/2$  and   maximum at $\theta=0$ and $\theta=\pi$, whereas that for the slow mode has two cut-offs at  $\theta=0$ and $\pi$,  and the maximum at $\theta=\pi/4$ and $3\pi/4$. 
 \par
 From Fig. \ref{fig1}, the growth rate can be estimated as $\gamma_{+}/\Omega_{0}\sim10^{-3}$ for $\Omega_0\sim10^4$ rad/s. Since $\Omega_{\nu+}\sim10^8$ rad/s for a small value of $k\sim10^2$,  we have  $\gamma_{+}/\Omega_{\nu+}\sim10^{-7}\ll1$, i.e.,  the weak beam and weak flavor oscillations assumptions hold for the fast mode in the entire regime of $\theta$. However, for the slow mode, the similar estimation applies except at $\theta=\pi/2$ where   $\Omega_{\nu-}=0$ and   $\gamma_{-}$ is not defined.    Thus, in difference to the fast magnetosonic  mode, which is   likely to be more unstable in the parallel and anti-parallel propagation, the slow mode   becomes unstable   for propagation at angles $\theta=\pi/4$ and $3\pi/4$.  Such distinctive features of the instability growth rates of magnetosonic modes  have not been reported in the earlier work  \cite{haas2017}.  
 \par 
 On the other hand,  when the magnetic field strength is relatively high, the similar features as in Ref. \cite{haas2017} (but in contrast to the previous case  with a low magnetic field) are noticed. The growth rate for the fast mode is found to be slightly increased by the influence of the flavor oscillations. This increase is, however, pronounced except for parallel and anti-parallel propagation. The weak perturbation assumptions, as said before, still hold for this mode and it displays  a stronger instability for propagation nearly perpendicular to the magnetic field. The effects of the flavor oscillations on the instability growth rates are rather significant for the slow magnetosonic mode over the entire domain of the propagation angle except at $\theta=\pi/2$ where   $\Omega_{\nu-}=0$ and $\gamma_{-}$ is not defined. However, close to this value of $\theta$, the growth rate for the slow mode tends to reach its minimum value, implying the stability of the slow magnetosonic mode therein. The instabilities are rather stronger for the parallel and anti-parallel propagation.  Furthermore,  the growth rates for the fast waves decrease with increasing values of the magnetic field, whereas the opposite trend occurs for the slow waves. 
From Fig. \ref{fig1}, it is also noted that the growth rates for both the fast and slow magnetosonic modes remain finite at $\theta\rightarrow0$ and $\theta\rightarrow\pi$ as the values of $\gamma_\nu^{\pm}$ and $\gamma_\text{osc}^{\pm}$ assume nonzero values therein  and that  $\gamma_{\pm}\sim10$ s$^{-1}$ (i.e., $1/\gamma_{\pm}\sim10^{-1}$ s), and it becomes higher (or $1/\gamma_{\pm}$ becomes lower) with the influence of flavor oscillations. This means that the  instability of the magnetosonic wave occurs faster in presence of the flavor oscillations than that due to its absence. Also, this typical time of instability is relatively shorter than the characteristic time scale of supernova explosion $\sim1-100$ s.   Thus, the neutrino flavor oscillations have a remarkable impact on the instability of  magnetosonic waves in neutrino-beam driven   magnetoplasmas. 
\par
\begin{figure*}[ht]
\centering
\includegraphics[height=2.8in,width=6.5in]{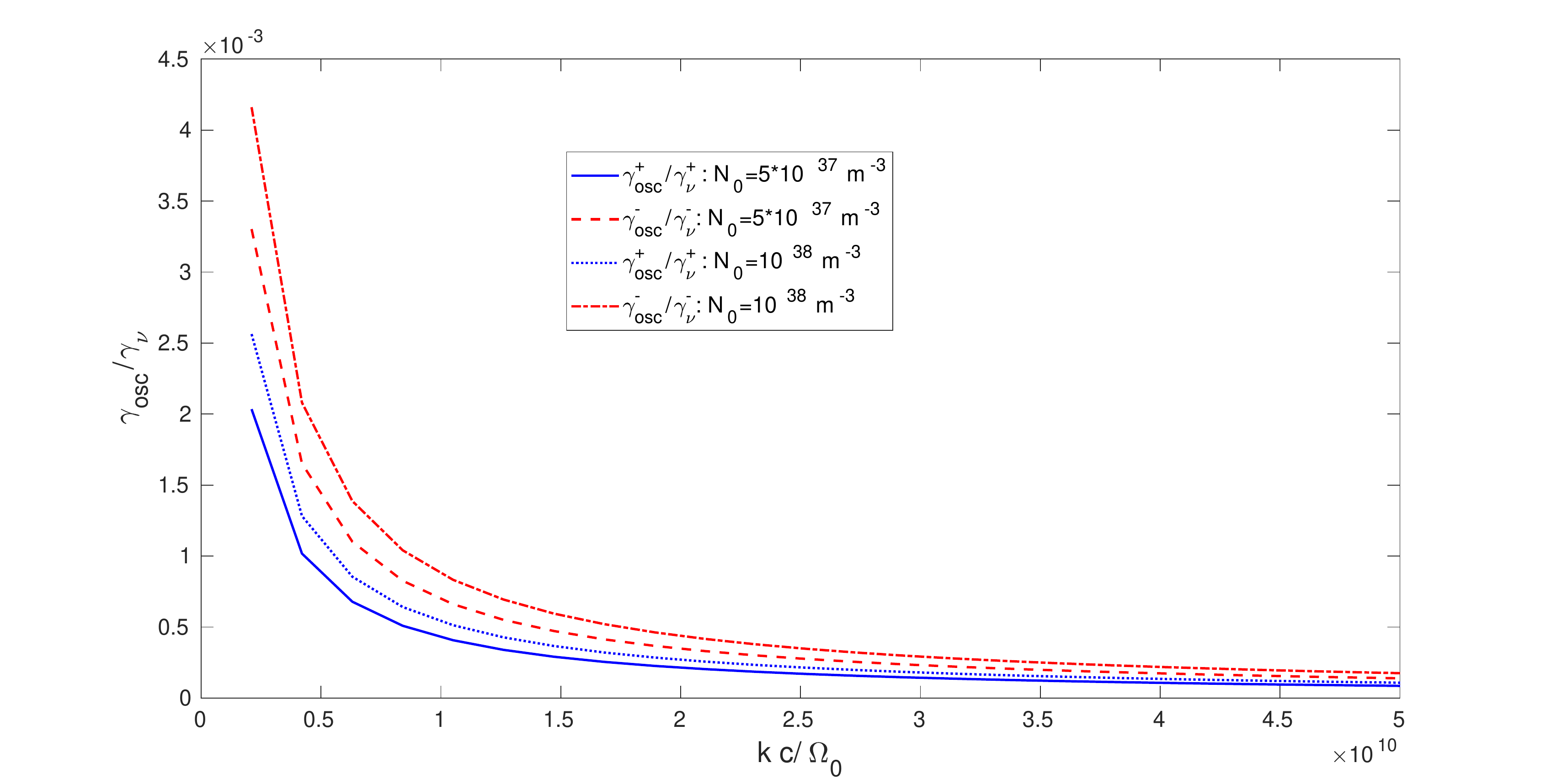}
\caption{The relative influences of the neutrino two-flavor oscillations $(\gamma^{\pm}_\text{osc})$ and the neutrino streaming beam $(\gamma^{\pm}_{\nu})$ on the growth rates of instabilities of oblique magnetosonic waves are shown against the wave number for two different values of the total neutrino density $N_0$ as in the legend. The fixed parameter values are $B_0=5\times10^6$ T, $\theta=\pi/4$ and others   as in the text.  }
\label{fig2}
\end{figure*}

 
The relative influence of the neutrino flavor oscillations can be qualitatively analyzed in two different density regimes and with the variations of the wave number. It is evident from Fig.  \ref{fig2}   that as one approaches towards a domain of higher wave numbers, the relative influence of flavor oscillations  on the instability growth rate becomes less important. However,   it can be  significant  for both the slow and fast magnetosonic modes at some higher density regimes, provided one   limits the neutrino density below a critical value, since at higher density regimes the relativistic degeneracy of electrons should come into the picture \cite{haas2019} which is not considered in the present model.  
   
 \section{Conclusion} \label{sec-conclu}
We have studied the influences of  the neutrino two-flavor oscillations  on the propagation of MHD waves and instabilities that are driven by the streaming neutrino beams. Special emphasis is given to analyzing the characteristics of fast and slow magnetosonic waves propagating in an arbitrary direction  to the static magnetic field.  In this way, the previous theory of neutrino MHD  waves    \cite{haas2017} is generalized with the inclusion of neutrino flavor oscillations in the wave dynamics. Using the neutrino MHD equations and assuming the weak neutrino beam-plasma interactions,   as well as the weak coupling between MHD waves and two-flavor oscillations, a general dispersion relation for MHD waves is derived which accounts for the contributions from the resonant-like interactions of MHD waves with both the streaming neutrino beam and flavor oscillations.  It is found that while the latter modify   the wave dispersion and  intensify the magnitude of the instability of oblique magnetosonic waves, the shear-Alfv{\'e}n wave remains uninfluenced by the neutrino beam and neutrino flavor oscillations. 
The growth rates of instabilities, so generated due to the energy exchange of neutrinos with the wave, are obtained and analyzed with the parameters relevant to type-II supernova explosion due to the core collapse of  massive stars. It is found that the relative influence of the flavor oscillations can be significant in the regimes of high neutrino beam densities and/or relatively low magnetic fields provided the wavelength is moderate or high.    Here, one must restrict the neutrino beam density to some limit. Otherwise,  in high density regimes, where   the relativistic degeneracy effects will come into the picture, one must deal with the relativistic NMHD model which is, however, a project for our future work. Furthermore, in the regimes of two different magnetic fields, the growth rate profiles exhibit almost opposite characters, implying that the instabilities can   be strong enough not only for the parallel and perpendicular propagation of waves but also for other directions of propagation with $\theta=\pi/4$ and $\theta=3\pi/4$.
\par 
To conclude, since the growth rate of instability  becomes higher (or its inverse becomes lower $\lesssim10^{-1}$ s) due to the effects of the neutrino flavor oscillations, the  instabilities of magnetosonic waves can occur within a shorter time than that in absence of the flavor oscillations. Consequently, this instability  should appear to be fast enough to provoke the neutrino radiation and two-flavor neutrino mixing in core-collapse supernovae \cite{muller2020}, since the typical time scale for supernova explosion is $\sim1-100$ s. Although the present model restricts to two neutrino flavor oscillations, the extension to three flavor states is also seemingly  important, however,  left for future studies. 
\section*{Acknowledgments}
   D. Chatterjee acknowledges support from Science and Engineering Research Board (SERB) for a
national postdoctoral fellowship (NPDF) with sanction order No. PDF/2020/002209 dated December 31, 2020.
\section*{Data availability statement}
The data that support the findings of this study are available upon reasonable request from the authors.

\bibliographystyle{apsrev4-1}
\bibliography{Reference}

\end{document}